\documentclass{icrc2009}

\usepackage{graphicx}   
\usepackage{fixltx2e}
\usepackage{url}

\newcommand{\shorttitle}[1]%
{\markboth{Proceedings of the 31\MakeLowercase{$^{st}$} ICRC, {\L}\'{o}d\'{z} 2009}{#1} }

\begin{document}
\title{Positioning system of the ANTARES Neutrino Telescope}

\author{\IEEEauthorblockN{Anthony M Brown\IEEEauthorrefmark{1}\IEEEauthorrefmark{2} on behalf of the ANTARES Collaboration \IEEEauthorrefmark{3}}
                            \\
\IEEEauthorblockA{\IEEEauthorrefmark{1}Centre de Physique des Particules de Marseille, 164 Av. de Luminy, Case 902, Marseille, France.}
\IEEEauthorblockA{\IEEEauthorrefmark{2}brown@cppm.in2p3.fr}
\IEEEauthorblockA{\IEEEauthorrefmark{3}\url{http://antares.in2p3.fr}}
}

\shorttitle{A.M. Brown - Antares Positioning System}
\maketitle

\begin{abstract}
 Completed in May 2008, the ANTARES neutrino telescope is located 40 km off the coast of Toulon, at a depth of about 2500 m. The telescope consists of 12 detector lines housing a total of 884 optical modules. Each line is anchored to the seabed and pulled taught by the buoyancy of the individual optical modules and a top buoy. Due to the fluid nature of the sea-water detecting medium and the flexible nature of the detector lines, the optical modules of the ANTARES telescope can suffer from  deviations of up to several meters from the vertical and as such, real time positioning is needed. 

Real time positioning of the ANTARES telescope is achieved by a combination of an acoustic positioning system and a lattice of tiltmeters and compasses. These independent and complementary systems are used to compute a global fit to each individual detector line, allowing us to construct a 3 dimensional picture of the ANTARES neutrino telescope with an accuracy of less than 10 cm. 

In this paper we describe the positioning system of the ANTARES neutrino telescope and discuss its performance during the first year of 12 line data taking.
  \end{abstract}

 
\section{Introduction}

Deployed off the coast of Toulon, the ANTARES telescope is, at present, the largest neutrino detector in the Northern hemisphere \cite{pc},\cite{large}. Utilising the Mediterranean Sea as a detecting medium, ANTARES consists of 12 detector lines, spaced approximately 70 meters apart, giving an overall surface area of the order of $~$0.1 km$^2$. The detection principle of ANTARES relies on the observation of Cherenkov photons emitted by charged relativistic leptons, produced through neutrino interactions with the surrounding water and seabed, using a 3 dimensional lattice of photomultiplier tubes (PMTs) \cite{pmt}. For ANTARES this 3 dimensional lattice of PMTs has been optimised for the detection of upward going high energy muon neutrinos and as such, each line of the detector, with the exception of Line 12, consists of 25 storeys separated by a vertical distance of 14.5 metres, each containing a triplet of Optical Modules (OMs) looking downwards at an angle of 45$^\circ$ from the vertical.

An important characteristic of any neutrino telescope is its angular resolution. At low energies the angular resolution is dominated by the angle between the parent neutrino and the resultant relativistic lepton. At larger energies, the angular resolution is dominated by the reconstruction of the relativistic lepton's track. Uncertainty in the reconstruction of the lepton's track is primarily governed by (i) the timing resolution of the individual OMs and (ii) the positional accuracy of where the OMs are located. 

The ANTARES expected angular resolution becomes better than 0.3$^\circ$ for neutrinos above 10 TeV in energy. To achieve this angular resolution, in-situ timing and positioning calibration are needed. ANTARES timing calibration is primarily achieved through the use of LED beacons deployed throughout the detector and is discussed elsewhere \cite{LED},\cite{LED2}. This paper describes the positioning calibration system of the ANTARES neutrino telescope and reviews its performance during the first year of 12 line operation. 
 
\section{ANTARES positioning calibration system.}
  
Each of the ANTARES 12 detector lines are anchored to a bottom string socket (BSS) on the seabed and pulled taught by the buoyancy of the individual OMs and a top buoy. Due to the flexible nature of these detector lines, even a relatively small water current velocity of 5 cm/s can result in the top storeys being displaced by several meters from the vertical. Therefore, real time positioning of each line is needed. This is achieved through two independent systems: an acoustic positioning system and a lattice of tiltmeters-compasses sensors. The shape of each line is reconstructed by performing a global $\chi^{2}$ fit using information from both of these systems. The relative positions of each individual OM is then calculated from this line fit using the known geometry of each individual storey. 

\begin{figure*}[th]
  \centering
  \includegraphics[width=6.0in]{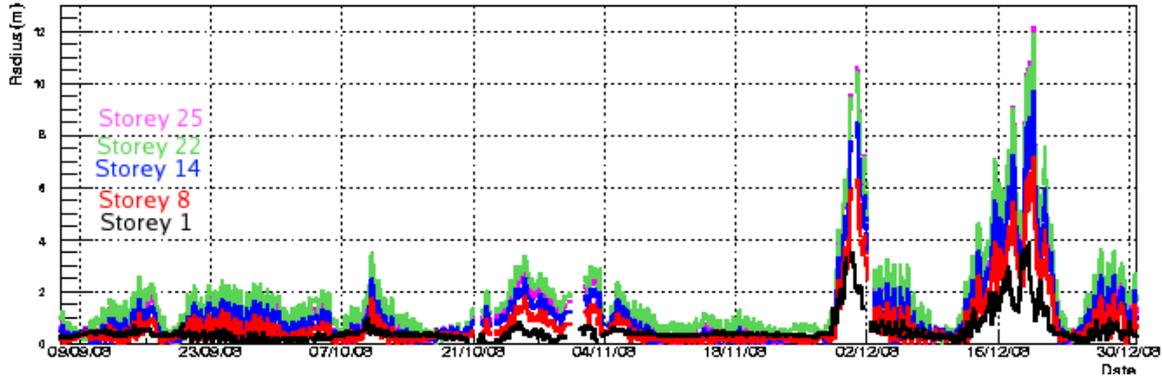}
  \caption{Radial displacement of all hydrophones on Line 11 during a 16 week period.}
  \label{radial}
 \end{figure*}

\subsection{Acoustic Positioning System}

The Acoustic Positioning System (APS) consists of a 3 dimensional array of emitters and receivers exchanging high frequency (40$-$60 kHz) acoustic signals. The use of high frequency signals allows a greater positional accuracy at the expense of a smaller acoustic attenuation length. Nonetheless, the 700$-$1000 meters attenuation length for 40$-$60 kHz signals is sufficient for the ANTARES telescope. The acoustic emitters are located on the BSS of each line, with an additional independent autonomous emitter being located approximately 145 meters from the array. 5 acoustic receivers are non-uniformily distributed along each line, with a higher density towards the top of the line where the line deflection due to the sea current is at its greatest. 

Measurements for the line shape reconstruction are performed every 2 minutes with the completion of an `acoustic run'. At present, each acoustic run consists of 14 cycles; each individual cycle being associated with the emission of an acoustic signal from a single acoustic emitter. For each cycle, the signal transit time to the different receivers is recorded. This information is used, in combination with the sound velocity profile, to calculate the distance between the acoustic emitter and the individual receivers. It should be noted that there are several sound velocity profilers located throughout the ANTARES detector, which measure the sound velocity profile at any given moment. These calculated distances are then used to triangulate the position of each acoustic receiver relative to the acoustic emitters. Figure \ref{radial} shows the radial displacement of all hydrophones on Line 11 during a 16 week period.

\subsection{Tiltmeter-Compass System}

The Tiltmeter-Compass System (TCS) gives the two perpendicular tilt angles, as well as the heading angle, for each storey. This information is given by a single TCM device, containing both tiltmeters and a compass, with a TCM device being installed in the electronics module of each storey. The range of measurement for the tiltmeters is $\pm 20^{\circ}$ on 2 perpendicular axis (roll and pitch), with an accuracy of $0.2^{\circ}$, while the compasses measures $360^{\circ}$ of heading at a resolution of $1^{\circ}$. Data is read out from the TCM device every 2 minutes and is used in conjunction with the APS during the line fit.  

\subsection{Line Shape Fit}

The shape of each detector line is reconstructed based upon a global $\chi^{2}$ fit to a line shape model. This model predicts the mechanical behaviour of the line under the influence of a sea current, taking into consideration the drag and buoyancy coefficients for each individual element along the detector line. At any given point, $i$, along the line, the zenith angle, $\theta_{i}$, from the BSS position, is given by:

   \begin{equation}
    tan(\theta_{i}) = \frac{\sum\limits_{j=i}^{N} F_{j}}{\sum\limits_{j=i}^{N} W_{j}}
    \label{theta1}
   \end{equation}
   
where $F_j$ is the flow resistance, or drag, and $W_j$ is the effective weight of the element in water, given by the weight of the element in air minus the buoyancy of the element. The flow resistance for each individual element on the line is calculated by Eq. \ref{fj}, where $\rho$ and $v$ is the fluid density and velocity respectively, $A_j$ is the cross-sectional area of the element and $C_w$ is the drag coefficient.

   \begin{equation}
    F_j = \frac{1}{2\rho C_{wj}A_jv^2}
    \label{fj}
   \end{equation}
   
Since $tan(\theta_i)$ is also the ratio of the radial displacement of the element from the vertical and the vertical displacement of the element from the base of the line, ($tan(\theta_i) = dr/dz$), then integrating along the line gives us the radial displacement as a function of altitude:

   \begin{equation}
    r(z)= av^2z - bv^2ln[1-cz]
    \label{rz}
   \end{equation} 

where $a$, $b$, and $c$ are known mechanical constants and $v$ is the sea current which is treated as a free parameter during the line fit. A global $\chi^{2}$ fit of this `line shape formula' is then performed on the combined data from the APS and the TCS. By doing so, a complete 3 dimensional positioning of the ANTARES neutrino telescope is performed every 2 minutes. It should be noted that, while a displacement of several meters can occur for the top storeys, the line movements are fairly slow, and therefore completing a full detector positioning every 2 minutes is more than adequate to achieve the 10 cm positional accuracy required for a good angular resolution of the apparatus.

\section{Performance}

Construction of the ANTARES neutrino telescope was completed in May 2008 with the deployment and connection of the last two detector lines. Throughout its construction, the ANTARES neutrino telescope has been operating with an increasing size. During these periods of data taken, the positioning system was extensively studied and proved to be stable in its operation (see \cite{keller}, \cite{vlv}). Here we will concentrate on the results of the positioning systems operation for the first year of data taking in the completed 12 line configuration. 



As mentioned, the global $\chi^{2}$  fit of the line shape formula considers data from both the APS and TCS. However, these two systems are independent, with the data from each system being analysed separately. Firstly, let us consider the contribution from the APS. 

\begin{figure}[!t]
  \centering
  \includegraphics[width=3.0in]{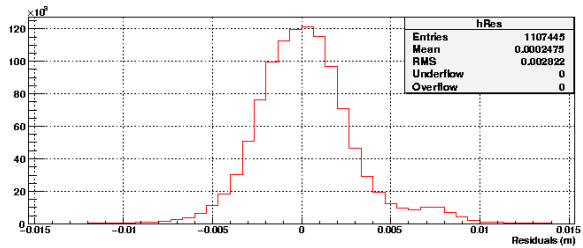}
  \caption{The difference between the distance calculated from the acoustic cycles and the distance re-calculated after triangulation, for Line 12, storey 1.}
  \label{apscompare}
 \end{figure}

In order to evaluate the stability of the triangulation of the line positions using the APS, the estimates of the distance between each pair of emitter and receiver, are compared before and after triangulation; the spread in this distribution being an indication of the accuracy of the triangulation. For the purpose of illustration, Figure \ref{apscompare} shows that the difference between these two estimated distances is of the order of millimeters and thus, considering acoustic data alone, reconstruction of the individual OMs can easily be achieved with a resolution better than the 10 cm required.  
 
\begin{figure}[!t]
  \centering
  \includegraphics[width=3.0in]{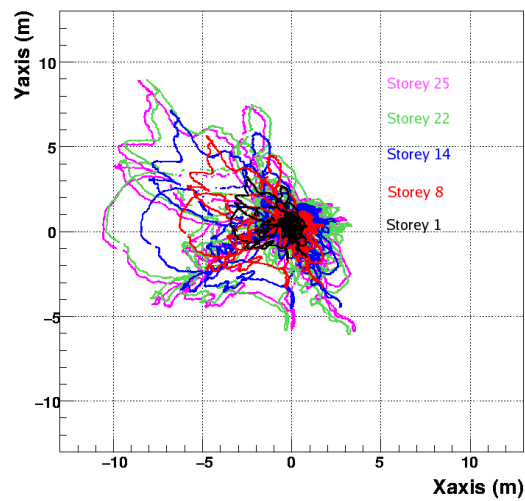}
  \caption{The horizontal movements from BSS, of all hydrophones on Line 11 for a 6 month period. The dominant East-West heading of the line movements is due to the dominant Ligurian current which flows at the ANTARES site.}
  \label{acou_tri}
 \end{figure}

The strength of the APS can be seen in Figure \ref{acou_tri}, which illustrates the horizontal movements of the line with respect to the BSS position, over a 6 month period. As illustrated in Figure \ref{acou_tri}, the top storeys of the detector line experience the largest amount of displacement due to the water current. Furthermore, this movement is mostly confined to an East-West heading, due to the dominant Ligurian current which flows at the ANTARES site. 
 
\begin{figure}[!t]
  \centering
  \includegraphics[width=3.0in]{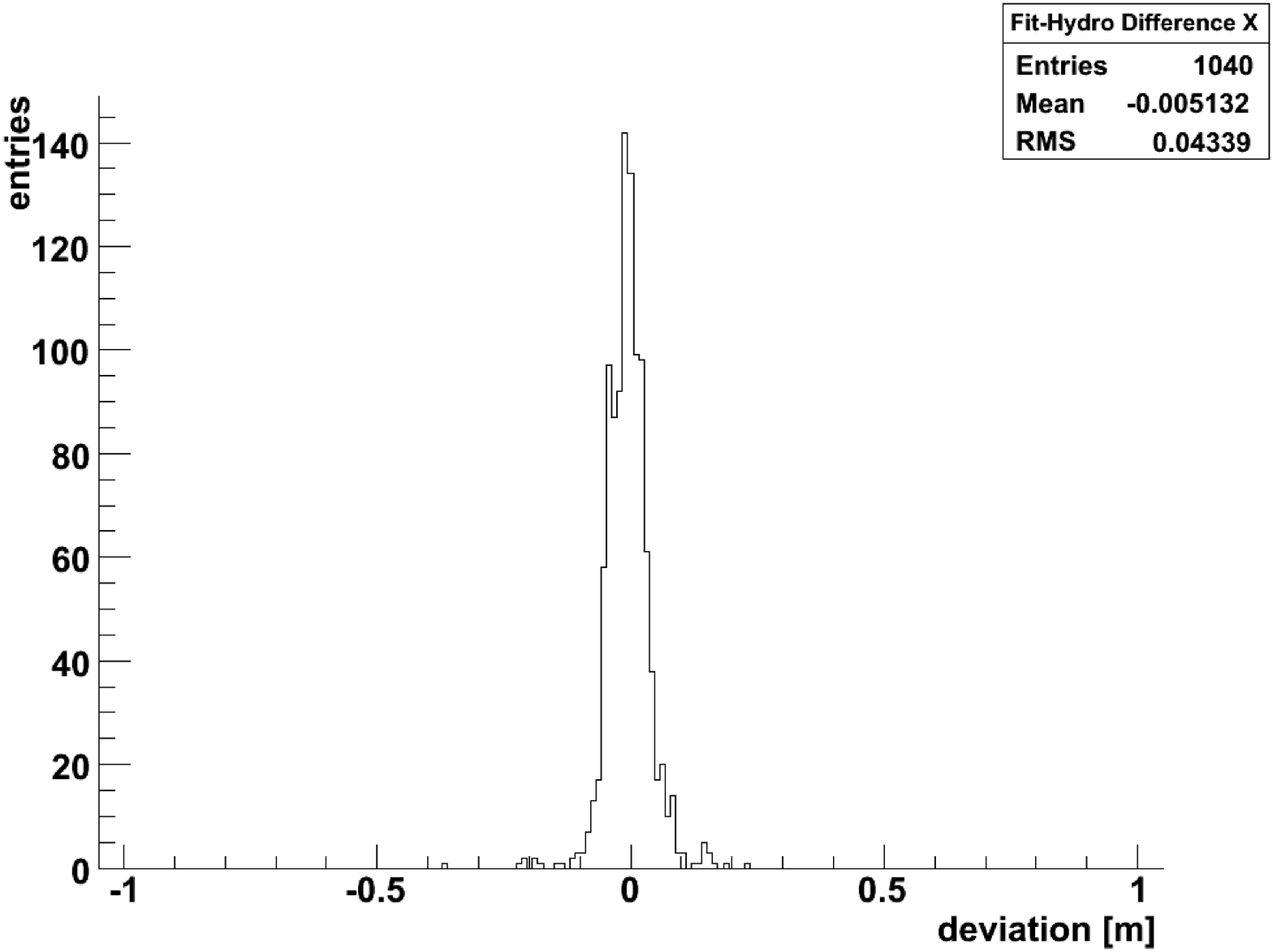}
  \caption{The difference between the X-position of storey 14 on Line 1 calculated by the hydrophone data and the X-position obtained by the line fit.}
  \label{compare}
 \end{figure}

To evaluate the compatibility of the data from the APS and TCS, in the overall line shape reconstruction, the line shape reconstructions, as based on the measurements of the APC and TCS alone, are compared. An illustration of such a comparison is shown in Figure \ref{compare}. As with Figure \ref{apscompare}, the spread in the distribution of this comparison is an indication of the compatibility of the two data sets, with a large spread implying a incompatibility of the data from the APS and TCS with regards to the overall line shape fit. As can be seen in Figure \ref{compare} the mean difference between the X position, as calculated using the two different data sets, is less than 1 cm. It should be noted that similar distributions are observed for the differences in the Y and Z positions of the hydrophones. 

\section{Conclusions}

ANTARES positioning system is stable and accurate in its operation for the complete 12 line configuration. Using the two independent systems of acoustic positioning and tiltmeters-compasses, we are able to reconstruct the 3 dimensional position of the OMs to an accuracy of 10 cm or less. Completing a 3 dimensional positioning of ANTARES every 2 minutes, the quasi-instantaneous knowledge of the line's position allows us to minimise the uncertainty in ANTARES angular resolution due to the fluid nature of our detecting medium. Furthermore, with its stability over a large detector volume, the current ANTARES positioning system is a sound starting point for the design of the positioning system for the future sea-based cubic kilometer sized neutrino telescope of KM3NET \cite{km3v2}.

 {\bf
 Acknowledgments}

Anthony Brown acknowledges the financial support of both the French funding agency `Centre National de la Recherche Scientifique' (CNRS) and the European KM3NET design study grant.

\end{document}